\documentclass[10pt, conference, letterpaper]{IEEEtran}

\usepackage[english]{babel}
\usepackage[usenames]{color}
\usepackage[cp1250]{inputenc}
\usepackage{amsfonts}
\usepackage{amsthm}
\usepackage{graphicx}
\usepackage{epsfig}
\usepackage{algorithm}
\usepackage{algorithmic}

\DeclareGraphicsExtensions{.jpg}
\newcommand{\be}{\begin{equation}}
\newcommand{\ee}{\end{equation}}

\theoremstyle{plain}
\newtheorem{df}{Definition}

\newtheorem{obs}[df]{Observation}

\begin{document}

\title{Practical estimation of rotation distance \\ and induced partial order for binary trees}
\author{\IEEEauthorblockN{Jarek Duda}
\IEEEauthorblockA{Jagiellonian University, Golebia 24, 31-007 Krakow, Poland. Email: \emph{dudajar@gmail.com}}}
\maketitle

\begin{abstract}
Tree rotations (left and right) are basic local deformations allowing to transform between two unlabeled binary trees of the same size. Hence, there is a natural problem of practically finding such transformation path with low number of rotations, the optimal minimal number is called the rotation distance. Such distance could be used for instance to quantify similarity between two trees for various machine learning problems, for example to compare hierarchical clusterings or arbitrarily chosen spanning trees of two graphs, like in SMILES notation popular for describing chemical molecules.

There will be presented inexpensive practical greedy algorithm for finding a short rotation path, optimality of which has still to be determined. It uses introduced partial order for binary trees of the same size: $t_1 \leq t_2$ iff $t_2$ can be obtained from $t_1$ by a sequence of only right rotations. Intuitively, the shortest rotation path should go through the least upper bound or the greatest lower bound for this partial order. The algorithm finds a path through candidates for both points in representation of binary tree as stack graph: describing evolution of content of stack while processing a formula described by a given binary tree. The article is accompanied with Mathematica implementation of all used procedures (Appendix).
\end{abstract}
\textbf{Keywords:} tree rotation, algorithmics, machine learning
\IEEEpeerreviewmaketitle
\section{Introduction}
Tree is a basic tool of computer science, used for example to represent hierarchical structure in data, e.g. in hierarchical clustering~\cite{hier}. It is a natural question to evaluate similarity between such trees, maybe propose a path of local deformations to get a transformation between such two structures. A natural candidate for the required elementary local deformation of a binary tree, maintaining order of leaves, are tree rotations (left and right), which switch levels of two nodes and reconnect their subtrees. It is used for example to balance binary search tree in popular AVL method~\cite{AVL}.

The minimal number of rotations to transform between two binary trees is a metric called rotation distance. It was shown that for any two $N$-node trees, for $n\geq 11$, this distance is at most $2N-6$ and there exist pairs of trees fulfilling this $2N-6$ distance~(\cite{rot1,rot2}). Binary trees have multiple equivalent representations, for example as bracketing or triangularization of polygon used in these two cited articles.
\begin{figure}[t!]
    \centering
        \includegraphics[width=8cm]{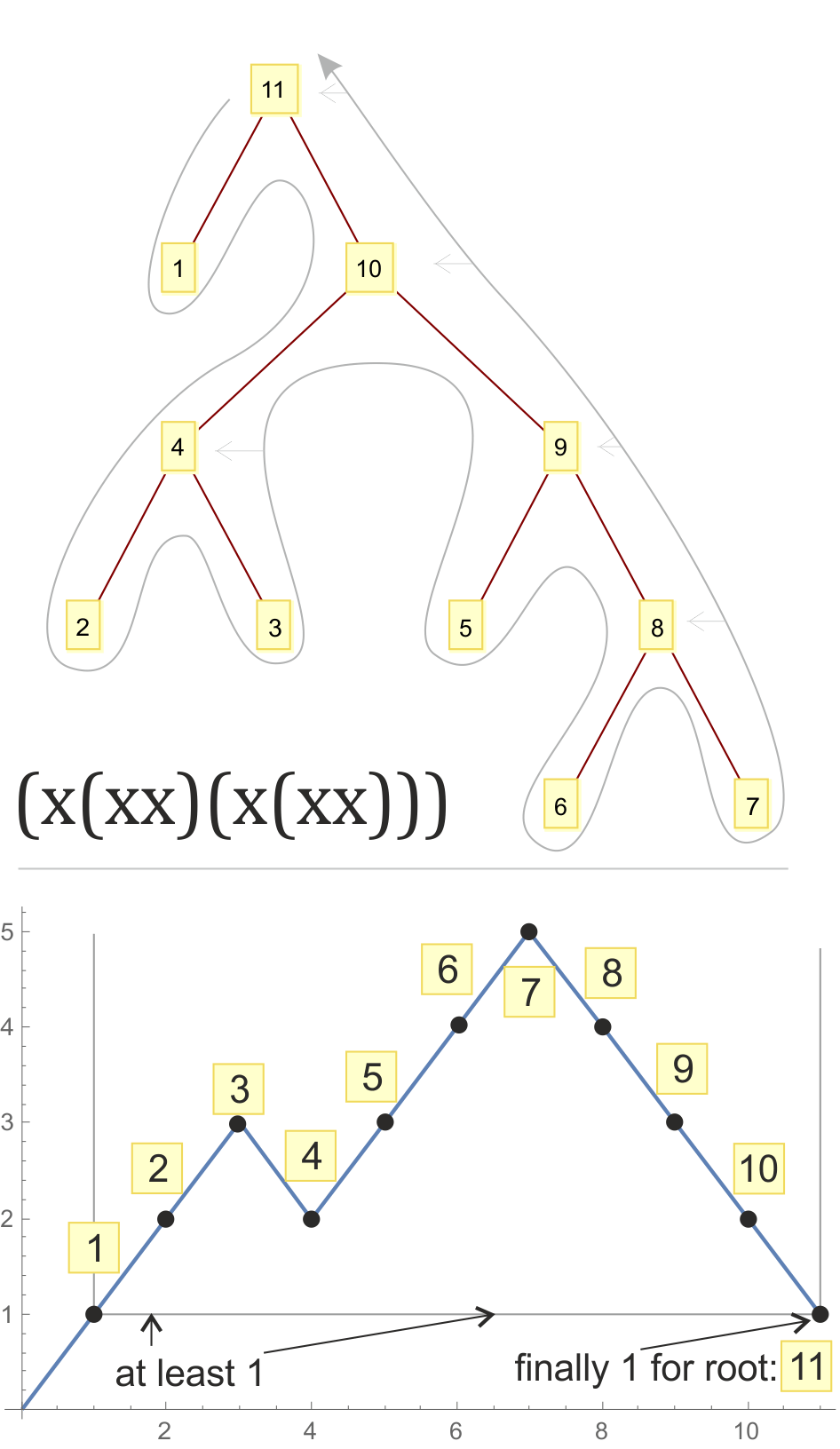}
        \caption{An example of binary tree (top), corresponding bracketing (center) and stack graph (bottom): describing evolution of content of stack while processing a formula described by a given binary tree (bracketing). It can be obtained by post-order tree traversal: visiting leaf corresponds to +1 for the graph ('x' in bracketing, push value to stack), visiting internal node to -1 (closing bracket: pop two values then push their function).}
        \label{stack}
\end{figure}

\begin{figure}[t!]
    \centering
        \includegraphics[width=8cm]{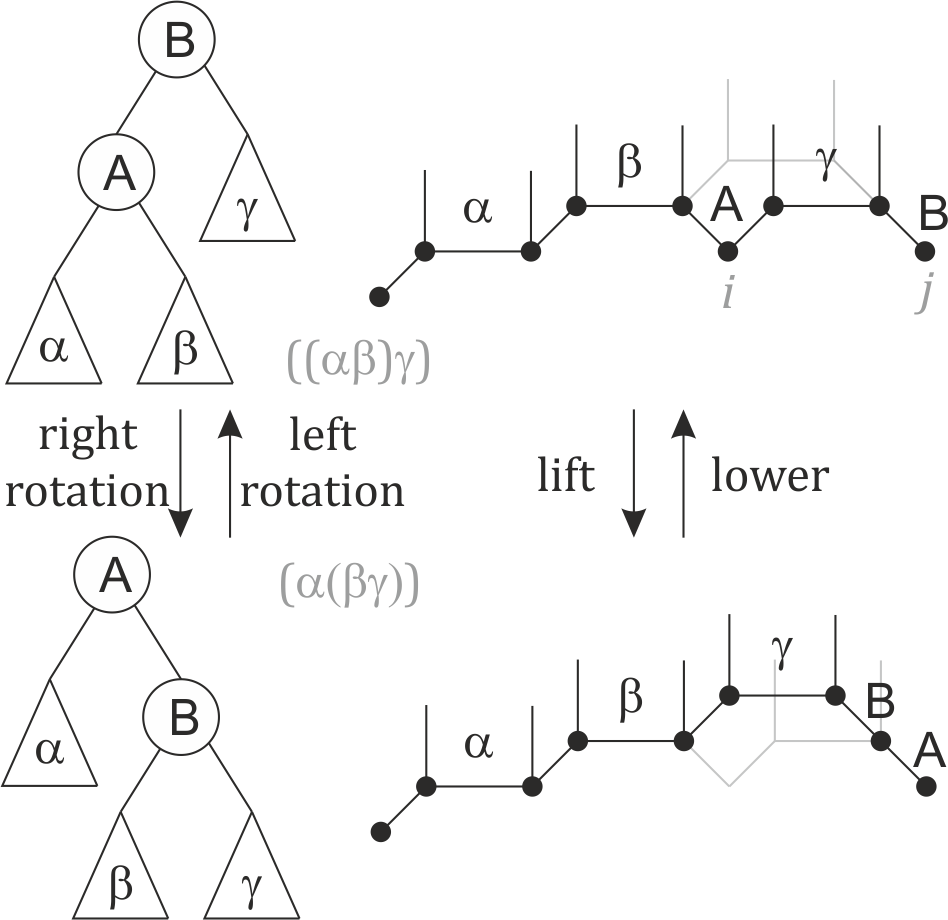}
        \caption{Tree rotations (left) and their analogue for corresponding bracketing (gray) and stack graph (right). The $\alpha$, $\beta$ and $\gamma$ are nonegative graphs of even width (can be zero), changing by $\pm 1$ per position. Presented lift will be denoted by $l_{ij}$, where $i,j$ are the marked positions.}
        \label{rot}
\end{figure}

There will be discussed greedy algorithm for finding a rotation path which base on a different representation: of evolution of stack content while processing such bracketed formula, which can be found for example in Chapter 7 of "Concrete mathematics" book~\cite{concrete}. We will refer to this function as \emph{stack graph}, it is visualized if Fig. \ref{stack} and left/right rotations become simple lower/lift procedures, visualized in Fig. \ref{rot}, which turn out convenient for searching for a short path.

\begin{figure}[t!]
    \centering
        \includegraphics[width=8cm]{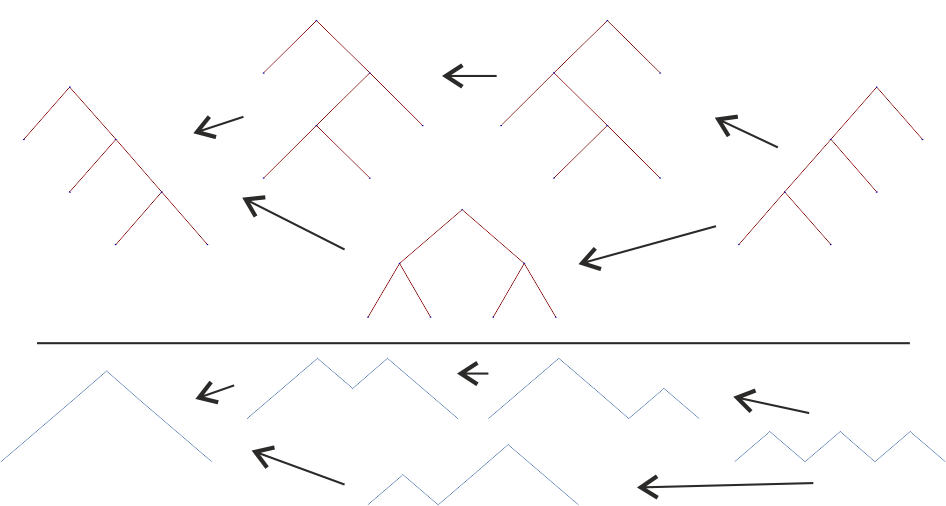}
        \caption{Partial order for all 4 leaf binary trees (top) and their stack graphs (bottom). Each arrow denotes possibility of single right rotation (or lift). }
        \label{vert4}
\end{figure}

\begin{figure*}[t!]
    \centering
        \includegraphics[width=16cm]{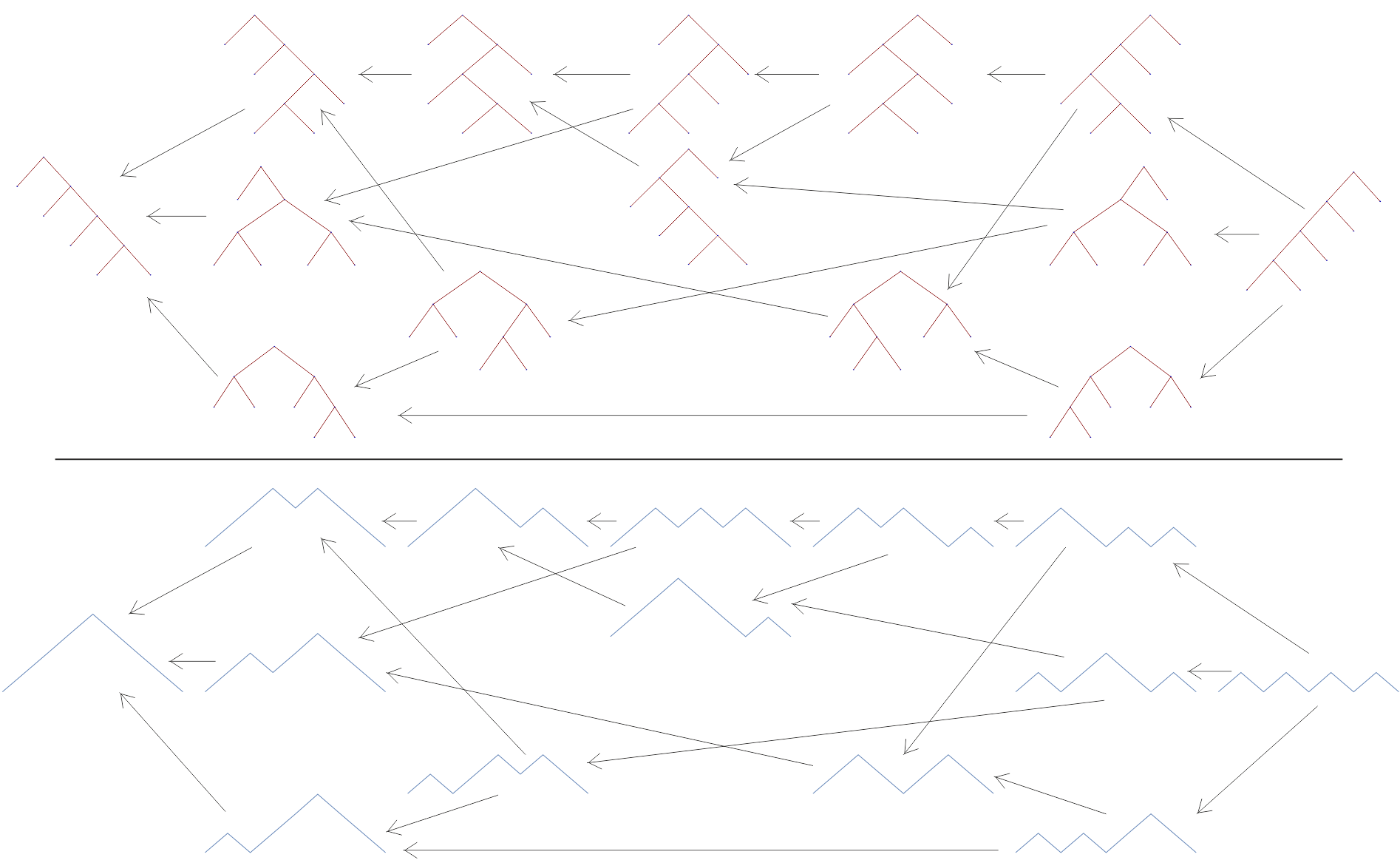}
        \caption{Partial order for all 5 leaf binary trees (top) and their stack graphs (bottom). Each arrow denotes possibility of single right rotation (or lift). Observe symmetry of the top diagram - taking mirror image of all trees switches left and right rotation, reversing the order}
        \label{vert5}
\end{figure*}

This representation allows to conclude that a natural relation is a partial order: $t_1 \leq t_2$ iff there exists a series of right rotations transforming tree $t_1$ to $t_2$. This partial order for 4 and 5 node trees is presented in Fig. \ref{vert4} and \ref{vert5}. These figures suggest that the shortest path between two trees has to go through their least upper bound ($\wedge$) or their greatest lower bound ($\vee$) for this order, or in other words: the path can be chosen (sorted) as first a sequence of right rotations, then of left rotations ($\wedge$) or oppositely ($\vee$). However, this intuition has still to be verified.

There will be presented a natural inexpensive ($O(n\cdot \textrm{'path length'})$) greedy algorithm to find the common lift (CL) for stack graphs of two trees, which corresponds to a candidate for the least upper bound ($\wedge$) of these trees. By taking mirror image of the trees, which switches left and right rotations and so reverses the order, this algorithm can also provide a candidate for the greatest lower bound ($\vee$). Its optimality has still to be verified, but currently it can be used to quickly find the upper bound for the rotation distance or to approximate this distance and find a short path, especially for situations where the optimality is not crucial.

For example for various machine learning situations, like evaluating distance between different hierarchical clusterings. Another example of situation where we need to quickly evaluate distance between trees is for comparing graphs for which we can define a spanning tree in an unique way. We cannot do it for general graphs, as it would solve the graph isomorphism problem, but such spanning trees are a popular tool to describe e.g. chemical molecules in so called SMILES notation~\cite{smiles}. Such trees are more complex than binary unlabeled: degree of nodes can be larger than two and both vertices and edges may have types worth including in the definition of distance - will require a generalization of the discussed method.

\section{Partial order and \\greedy search for common lift}
Denote the set of all $n$-leaf unlabeled binary trees as $T_n$, $|T_n|={2n-2 \choose n-1}/n$ is Catalan number. Each of such trees has $n-1$ internal nodes of degree 2, denote the number of all its nodes as $N=2n-1$.

Let us define its \emph{stack graph} as in Fig. \ref{stack}:
\begin{df}
Stack graph $s_t:\{0, \ldots, N\} \to \mathbb{N}$ for tree $t\in T_n$ is function defined by the following conditions:
\begin{itemize}
  \item $s_t(0)=0,\ s_t(N)=1$,
  \item for $i=1,\ldots,N$, $s_t(i)-s_t(i-1)=1$ if $i$-th node in post-order traversal of $t$ is leaf, $-1$ otherwise
\end{itemize}
\end{df}

Stack graph is fixed for 0 and 1, and is at least 1 for $i\in \{1,\ldots,N\}$. For better visualization, it will be depictured with points joined by lines, sometimes without the fixed $[0,1]$ range (Fig. \ref{vert4}, \ref{vert5}). Denote $S_n=s(T_n)$ as the set of all stack graphs for $n$-leaf binary trees.

The tree rotation operations are presented in Fig. \ref{rot}. In bracketing notation, right rotation changes $((\alpha\beta)\gamma)$ into $(\alpha(\beta\gamma))$ and left rotation the opposite, where $\alpha,\ \beta, \gamma$ represent some formulas (subtrees), which can be degenerated into single variables (leaves).

As we can see in this figure, right rotation corresponds to lift by $(-1,1)$ vector of some $\{i,\ldots,j\}$ range. It can be degenerated: $i=j$ for leaves.
\begin{df}
For two stack graphs: $s_1, s_2\in S_n$ we will say that $s_2$ is $(i,j)$ lift of $s_1$, denoted as $s_2=l_{ij}(s_1)$, if the following conditions are fulfilled:
\begin{enumerate}
  \item $i \leq j$, $s_1(i)=s_1(j)$
  \item $s_1(i-1)=s_1(i+1)=s_1(j-1)=s(i)+1$
  \item for $i<k<j$, $s_1(k)>s_1(i)$
  \item for $k<i$ and $j<k$, $s_1(k)=s_2(k)$
  \item for $i\leq k\leq j$, $s_2(k-1)=s_1(k)+1$
\end{enumerate}
\end{df}
The value of stack graph will be referred as \emph{level}. Condition 1) says that $i\leq j$ are on the same level of $s_1$, 2) says that there is first '$\vee$' shape in $i$, then '$\backslash$' shape in $j$. Condition 3) enforces $s_1$ being above the $i,j$ level. Finally 4) and 5) say that $s_1$ and $s_2$ differ only on the $[i,j]$ range, in which $s_2$ is lift by $(-1,1)$ vector of $s_1$.

Obviously, lifting cannot decrease value in any position:
\begin{obs}
  If $t_2$ is a right rotation of $t_1$, then for all $i\in\{0,\ldots,N\}$, $s_{t_1}(i)\leq s_{t_2}(i)$.
\end{obs}
It allows to conclude antisymmetry of the following relation, its reflexivity and transitivity are obvious, making it a partial order in $T_n$:
\begin{df}
  For $t_1,t_2\in T_n$ we define partial order: $t_1\leq t_2$ iff $t_1=t_2$ or there exists a sequence of right rotations from $t_1$ to $t_2$.
\end{df}
Examples of this partial order for 4 and 5 leaf trees are presented in Fig. \ref{vert4} and \ref{vert5} correspondingly. These examples suggest that the shortest rotation path has to go through the least upper bound ($\wedge$) or the greatest lower bound ($\vee$), however,  it does not have to be generally true. Having algorithm searching for a short path going through the least upper bound ($\wedge$), we could use it to find path going through the greatest lower bound ($\vee$) by taking mirror images of both trees, which switch left and right rotations, reversing this order. There is a nontrivial relation between stack graph of a tree and of its mirror image, which examples are presented in various figures of this paper, but it can found in $O(n)$ time and memory (see \verb"mirror[]" in Appendix).

\begin{figure}[t!]
    \centering
        \includegraphics[width=8cm]{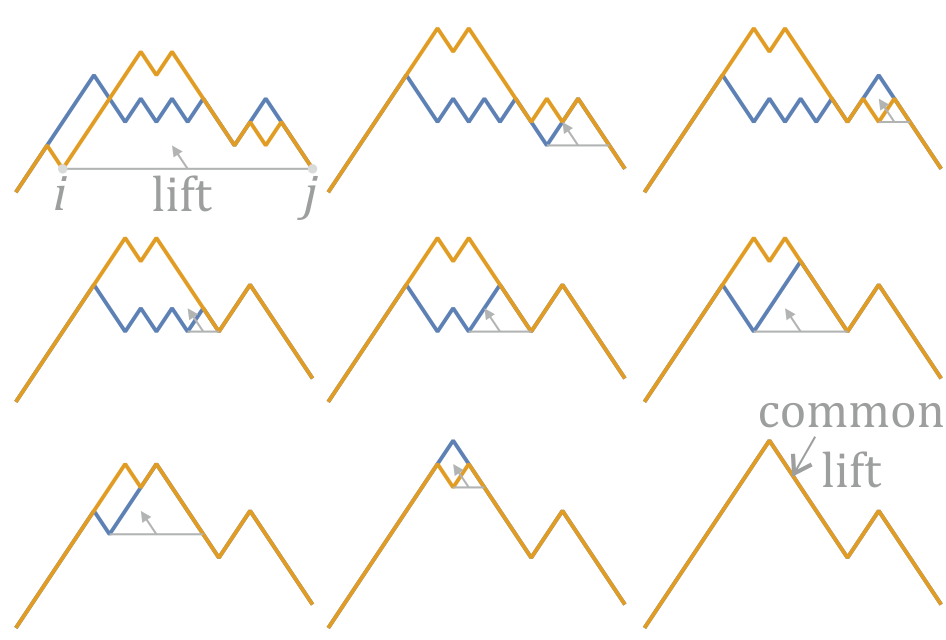}
        \caption{Example of finding a common lift using the greedy algorithm. Its step finds the lowest level where the two stack graphs disagree, find the most-right position of such points in this level ($i$), find the closest position in this level right to this point ($j$), then lift $(i,j)$ segment. Such steps continue until equalizing both stack graphs, getting a common lift. }
        \label{greedy}
\end{figure}

\begin{figure*}[b]
    \centering
        \includegraphics[width=18cm]{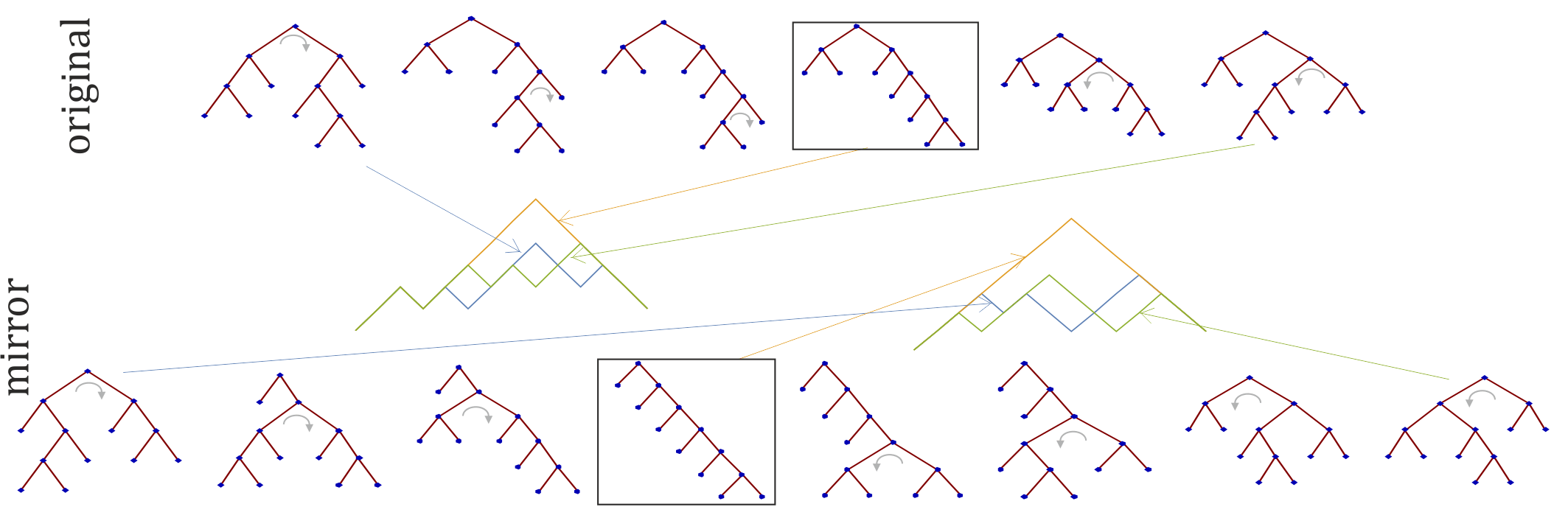}
        \caption{Example of rotation path found by the greedy algorithm for common lift, for the original trees (top) and their mirror images (bottom). The marked trees correspond to the common lift. The upper path is shorter so it will be the final answer. }
        \label{shortest}
\end{figure*}

To find a candidate of the the greatest lower bound, we can take both stack graphs and use a sequence of lifts applied to a locally lower one, to finally equalize them, getting a candidate for the lowest common lift and of path to them. Example of such process is presented in Fig. \ref{greedy}. It is presented (without tracking the changes) as Algorithm \ref{greed1} and \ref{greed2}.

\begin{algorithm}[htbp]
\caption{Greedy common lift for s1, s2 stack graphs}
\label{greed1}
\begin{algorithmic}
\WHILE {s1 $\neq$ s2}
\STATE lvl = min\{s1[k] : s1[k] $\neq$ s2[k]\}
\STATE i = max\{k : s1[k] $\neq$ s2[k], lvl = min(s1[k], s2[k])\}
\IF{s1[i] $<$ s2[i]}
\STATE j = min\{k $>$ i: s1[k] = lvl\}; lift(s1,i,j)
\ELSE
\STATE j = min\{k $>$ i: s2[k] = lvl\}; lift(s2,i,j)
\ENDIF
\ENDWHILE
\end{algorithmic}
\end{algorithm}

\begin{algorithm}[htbp]
\caption{Greedy shortest path for s1, s2}
\label{greed2}
\begin{algorithmic}
\STATE find common lift for s1 and s2
\STATE find common lift for mirror(s1) and mirror(s2)
\STATE take the shorter one
\end{algorithmic}
\end{algorithm}

\begin{figure*}[t]
    \centering
        \includegraphics[width=18cm]{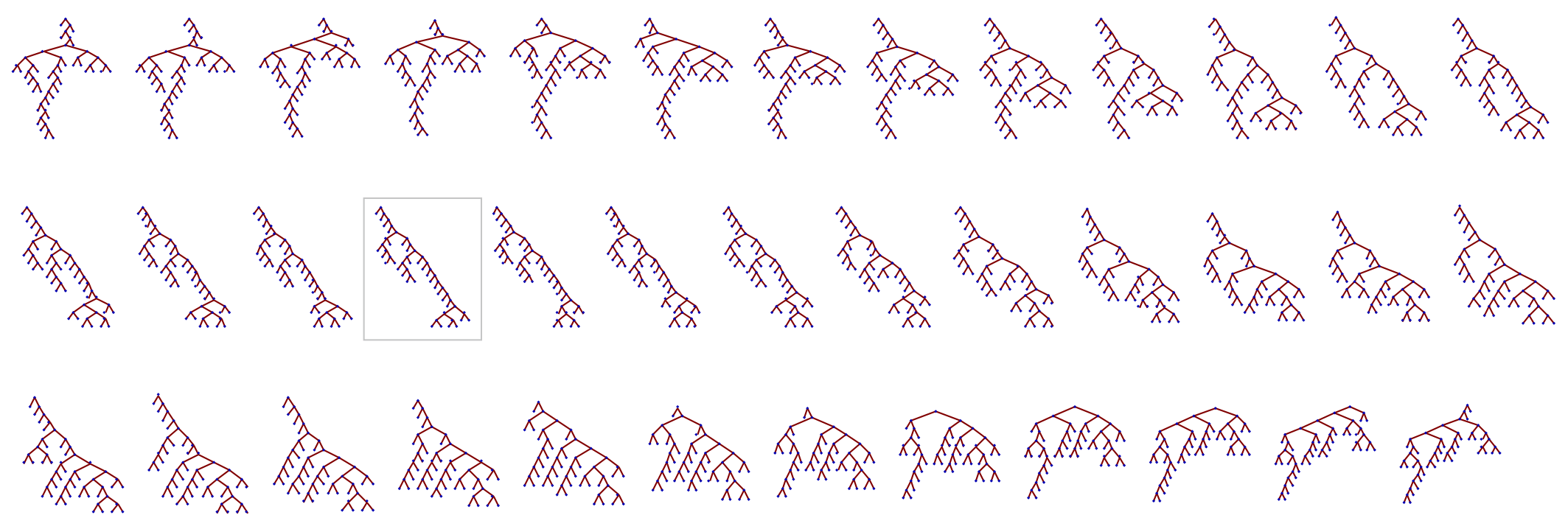}
        \caption{Example of found rotation path for two random 30 vertex trees. The marked tree corresponds to the common lift. }
        \label{30vertex}
\end{figure*}

Its naive implementation (see  \verb"findrotationpath[]" in Appendix) is linear in $n$ and path length. Figure \ref{clift} shows examples of common lifts found by this algorithm, Figure \ref{30vertex} shows example of found rotation path for tree with 30 leaves.

\begin{figure*}[t]
    \centering
        \includegraphics[width=18cm]{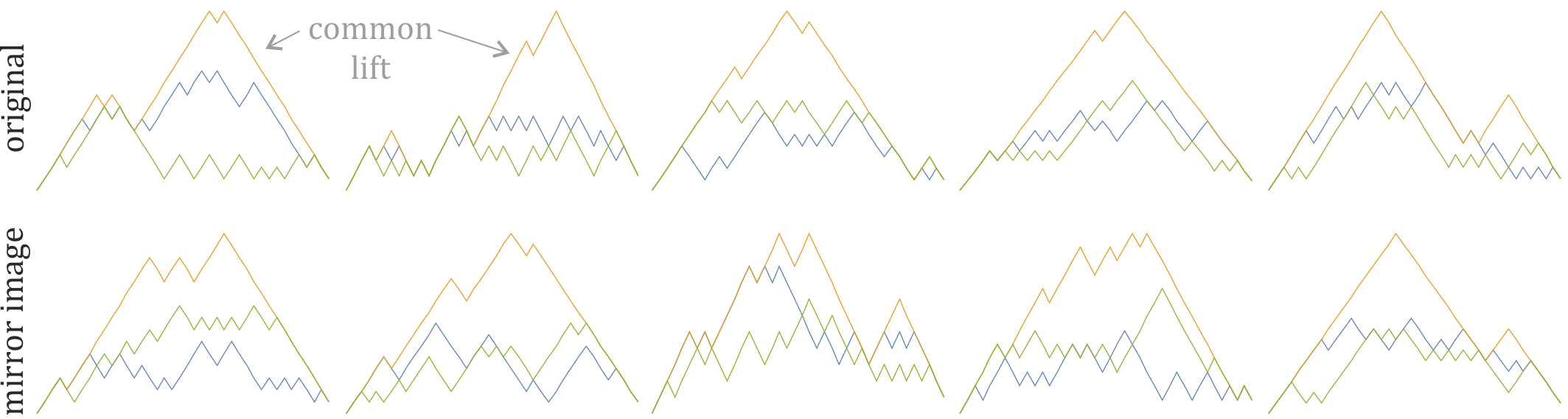}
        \caption{Five examples (columns) of the found common lift by greedy algorithm (highest orange line) for stack graphs of two random trees (lower: blue and green lines). The upper row corresponds to stack graphs of the original trees: common lift is a candidate for the least upper bound. The lower row corresponds to stack graphs of mirror images of the trees.}
        \label{clift}
\end{figure*}

\section{Conclusions and further perspectives}
There was presented a practical way to search for a short rotation path between two unlabeled binary trees of the same size. Its optimality is not guaranteed in this moment, nor counterexamples were found. It can be used to find a rotation path and bound from above or approximate the rotation path.

The basic question to investigate is optimality of this algorithm. If it is not optimal, maybe find its improvements, find bounds for its inaccuracy, try to characterise counterexamples. If it is already optimal, the proof might go through the following steps, each of which might turn out false:
\begin{enumerate}
  \item The shortest rotation path has to go through the least upper bound or the greatest lower bound of the two trees. In other words, the shortest rotation path can be chosen (sorted) as only right rotations first then only left rotations, or oppositely,
  \item The common lift found by the greedy algorithm corresponds to the least upper bound,
  \item The greedy algorithm finds the shortest path to the obtained common lift - the number of lifts cannot be reduced.
\end{enumerate}
Another large topic is applying this or extended methods for various problems, especially in machine learning, where not being optimal seems not crucial. There will be needed some expansions, like generalization to non-binary trees, which can be realized for example by splitting a higher degree node into a few binary nodes. Also, in some applications we should distinguish types of nodes and edges (e.g. atoms and molecular bonds), which might require modifying the metric and so the optimization problem.

A related line of work is trying to apply this method to evaluate similarity of graphs by comparing some their arbitrarily chosen spanning trees. Choosing an isomorphism independent spanning tree is generally a difficult problem, as it would allow to solve the graph isomorphism problem. However, it can be useful for some families of graphs, used for example in SMILES representation of chemical molecules.

\appendix
This appendix contains Wolfram Mathematica sources used to generate figures in this article. It generates random binary tree assuming uniform probability distribution among trees of given number of leaves. For this purpose, it uses enumerative decoding~\cite{enum} to get a random sequence of +1 and -1 which sums to 1, then finds its only cyclic rotation giving a stack graph.
\begin{scriptsize}
\begin{verbatim}
(* enumerative decoding *)
dec[n_, w_, X_] := (l = w; x = X;
Table[b = Binomial[n - i, l];
If[x < b, 0, x -= b; l--; 1], {i, n}]);
(* generate stack graph of n leaf random tree *)
randtree[n_] := (nn = 2 n - 1;
s = 2 dec[2 n - 1, n,
 RandomInteger[Binomial[2 n - 1, n - 1] - 1]] - 1;
Do[s[[i]] += s[[i - 1]], {i, 2, nn}];
min = Min[s];
div = Max[Table[If[s[[i]] == min, i, 0], {i, nn}]];
Join[s[[div ;; nn]], 1 + s[[1 ;; div]]] - min);

(* draw tree from stack graph *)
drawtree[s_] := (st = Table[0, {i, Length[s]}];
stp = 1; vn = 1; edg = {};
Do[If[s[[i + 1]] - s[[i]] == 1, st[[stp++]] = vn++,
 AppendTo[edg,
  {st[[stp - 2]] -> vn, st[[stp - 1]] -> vn}];
 st[[stp - 2]] = vn++; stp--], {i, 1, Length[s] - 1}];
 TreePlot[Sort[Flatten[edg]], Automatic, Length[s] - 1,
 VertexLabeling -> False]);

(* mirror image of tree from stack graph *)
mirror[s_] := (ls = Length[s]; p = 0;
pbr = Table[i, {i, ls/2}]; nbr = Table[0, {i, ls/2}];
Do[If[s[[i + 1]] - s[[i]] == 1, p++,
 brac = pbr[[pbr[[p]] - 1]];
 nbr[[pbr[[p]] = brac]]++], {i, ls - 1}];
S = Table[0, {i, ls}]; p = 1;
Do[p++; S[[p]] = S[[p - 1]] + 1;
 Do[p++; S[[p]] = S[[p - 1]] - 1, {j, 1, nbr[[i]]}]
 ,{i, ls/2, 1, -1}]; S);

(* the greedy algorithm to find path *)
lift[s_, f_, t_] :=
 Join[s[[1 ;; f - 1]], s[[f + 1 ;; t]] + 1, {s[[t]]},
  s[[t + 1 ;; Length[s]]]];
findrotationpath[S1_, S2_] := (
s1 = S1; s2 = S2; ll = {s1}; rl = {s2};
While[s1 != s2, min = Infinity;
 Do[If[s1[[i]] != s2[[i]], m = Min[s1[[i]], s2[[i]]];
  If[m <= min, min = m; pm = i]], {i, Length[s1]}];
 px = pm + 2;
 If[s1[[pm]] < s2[[pm]],
  While[s1[[px]] > min, px++];
  AppendTo[ll, s1 = lift[s1, pm, px]],
  While[s2[[px]] > min, px++];
  PrependTo[rl, s2 = lift[s2, pm, px]]]];
Join[ll, rl[[2 ;; Length[rl]]]])

(* example of application *)
n = 10; S1 = randtree[n]; S2 = randtree[n];
path1 = findrotationpath[S1, S2];
path2 = findrotationpath[mirror[S1], mirror[S2]];
Print[Row[Table[drawtree[path1[[i]]], {i, Length[path1]}]]]
Row[Table[drawtree[path2[[i]]], {i, Length[path2]}]]
\end{verbatim}
\end{scriptsize}

\bibliographystyle{IEEEtran}
\bibliography{cites}

\begin{thebibliography}{1}
\providecommand{\url}[1]{#1}
\csname url@samestyle\endcsname
\providecommand{\newblock}{\relax}
\providecommand{\bibinfo}[2]{#2}
\providecommand{\BIBentrySTDinterwordspacing}{\spaceskip=0pt\relax}
\providecommand{\BIBentryALTinterwordstretchfactor}{4}
\providecommand{\BIBentryALTinterwordspacing}{\spaceskip=\fontdimen2\font plus
\BIBentryALTinterwordstretchfactor\fontdimen3\font minus
  \fontdimen4\font\relax}
\providecommand{\BIBforeignlanguage}[2]{{%
\expandafter\ifx\csname l@#1\endcsname\relax
\typeout{** WARNING: IEEEtran.bst: No hyphenation pattern has been}%
\typeout{** loaded for the language `#1'. Using the pattern for}%
\typeout{** the default language instead.}%
\else
\language=\csname l@#1\endcsname
\fi
#2}}
\providecommand{\BIBdecl}{\relax}
\BIBdecl

\bibitem{hier}
J.~H. Ward~Jr, ``Hierarchical grouping to optimize an objective function,''
  \emph{Journal of the American statistical association}, vol.~58, no. 301, pp.
  236--244, 1963.

\bibitem{AVL}
G.~Adelson-Velsky and E.~Landis, ``An algorithm for the organization of
  information,'' \emph{Proceedings of the USSR Academy of Sciences}, vol. 146,
  pp. 263--266, 1962.

\bibitem{rot1}
D.~D. Sleator, R.~E. Tarjan, and W.~P. Thurston, ``Rotation distance,
  triangulations, and hyperbolic geometry,'' \emph{Journal of the American
  Mathematical Society}, vol.~1, no.~3, pp. 647--681, 1988.

\bibitem{rot2}
L.~Pournin, ``The diameter of associahedra,'' \emph{Advances in Mathematics},
  vol. 259, pp. 13--42, 2014.

\bibitem{concrete}
D.~E. Knuth, R.~L. Graham, O.~Patashnik \emph{et~al.}, ``Concrete
  mathematics,'' \emph{Adison Wesley,}, 1989.

\bibitem{smiles}
D.~Weininger, ``Smiles, a chemical language and information system. 1.
  introduction to methodology and encoding rules,'' \emph{Journal of chemical
  information and computer sciences}, vol.~28, no.~1, pp. 31--36, 1988.

\bibitem{enum}
L.~Oktem, \emph{Hierarchical enumerative coding and its applications in image
  compression}.\hskip 1em plus 0.5em minus 0.4em\relax Tampere University of
  Technology, 1999.

\end{thebibliography}
\end{document}